\documentclass[preprint,journal]{vgtc2}       %

\ifpdf%
\pdfoutput=1\relax                   %
\usepackage{graphicx}                %
\DeclareGraphicsExtensions{.pdf,.png,.jpg,.jpeg} %
\else%
\usepackage{graphicx}                %
\DeclareGraphicsExtensions{.eps}     %
\fi%

\graphicspath{{figures/}{pictures/}{images/}{./}} %

\usepackage{complexity}
\usepackage{amsmath}
\usepackage{todonotes}
\usepackage{tabularx}

\usepackage{microtype}                 %
\PassOptionsToPackage{warn}{textcomp}  %
\usepackage{textcomp}                  %
\usepackage{mathptmx}                  %
\usepackage{times}                     %
\renewcommand*\ttdefault{txtt}         %
\usepackage{cite}                      %
\usepackage{tabu}                      %
\usepackage{booktabs}                  %

\newcommand{\anais}[1]{\todo[color=red!50]{A: #1}}
\newcommand{\anaisinline}[1]{\todo[inline,color=red!50]{A: #1}}
\newcommand{\michael}[1]{\todo[color=yellow!50]{MH: #1}}
\newcommand{\michaelinline}[1]{\todo[inline,color=yellow!50]{MH: #1}}
\newcommand{\martin}[1]{\todo[color=blue!30]{MN: #1}}
\newcommand{\martinline}[1]{\todo[inline,color=blue!30]{MN: #1}}

\onlineid{7465}

\vgtccategory{Research}
\vgtcpapertype{algorithm/technique}

\title{MySemCloud: Semantic-aware Word Cloud Editing}

\author{Michael Huber, Martin Nöllenburg, and Anaïs Villedieu}
\authorfooter{
	\item
	Michael Huber, Martin Nöllenburg and Anaïs Villedieu are with the Algorithms and Complexity Group at TU Wien. E-mail: \{michael.huber|martin.noellenburg|anais.villedieu\}@tuwien.ac.at.
}
\shortauthortitle{Huber \MakeLowercase{\textit{et al.}}: MySemCloud: Semantic-aware Word Cloud Editing}
\abstract{Word clouds are a popular text visualization technique that summarize an input text by displaying its most important words in a compact image. The traditional layout methods do not take proximity effects between words into account; this has been improved in semantic word clouds, where relative word placement is controlled by edges in a word similarity graph.
	We introduce MySemCloud, a new human-in-the-loop tool to visualize and edit semantic word clouds.
	MySemCloud lets users perform computer-assisted local moves of words, which improve or at least retain the semantic quality.
	To achieve this, we construct a word similarity graph on which a system of forces is applied to generate a compact initial layout with good semantic quality. 
	The force system also allows us to maintain these attributes after each user interaction, as well as preserve the user's mental map.
	The tool provides algorithmic support for the editing operations to help the user enhance the semantic quality of the visualization, while adjusting it to their personal preference.
	We show that MySemCloud provides high user satisfaction as well as permits users to create layouts of higher quality than state-of-the-art semantic word cloud generation tools.} %


\CCScatlist{ %
	\CCScat{K.6.1}{Management of Computing and Information Systems}%
	{Project and People Management}{Life Cycle};
	\CCScat{K.7.m}{The Computing Profession}{Miscellaneous}{Ethics}
}

\teaser{
	\centering
	\includegraphics[width=\linewidth]{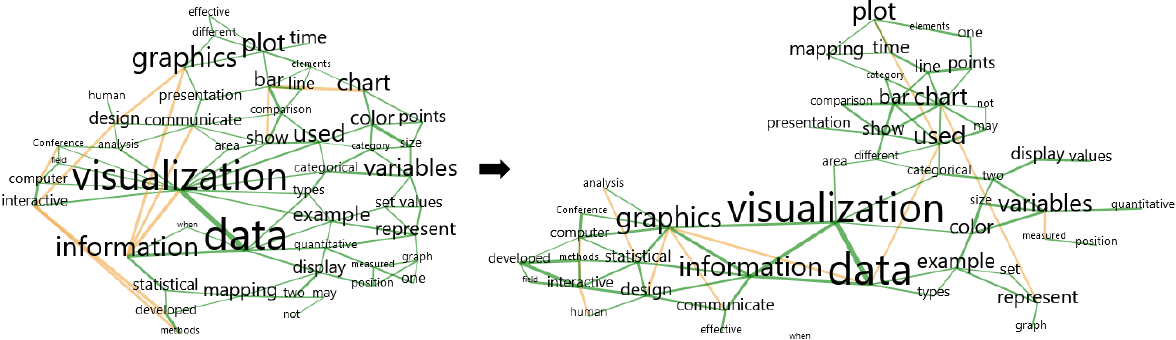}
	\caption{Semantic word cloud initially generated by our system from the ``data and information visualization'' Wikipedia page on the left. Realized adjacencies are shown in green and the main missing adjacencies in yellow. On the right the same word cloud after local user improvements. More adjacencies are realized, remaining large missing edges are shorter, and we can see three clusters emerge.}
	\label{fig:teaser}
}

\vgtcinsertpkg

\begin{document}

	\maketitle
	
	\section{Introduction}
	
	Word clouds are a common text visualization technique, where an input document is summarized into a compact visualization of its most important words, and the font size of each word is scaled proportionally to its frequency in the input text. Word clouds have gained popularity~\cite{ViegasWF09} through the automated tool Wordle~\cite{feinberg_2009} introduced in 2009. 
	While these colorful, compact layouts are aesthetically pleasing and playful, word clouds have been found to be a poor analytical information visualizations technique. When presented with a traditional layout that is purely optimized for compactness, users fail to properly identify the themes of the source text~\cite{HearstPPLLF20}. Instead, a proposed improvement is to use the positioning of the words in space to encode higher-level information about the text, e.g., by clustering together words that correspond to a shared overarching theme.
	
	Cui et al.~\cite{CuiWLWZQ10} proposed such a semantically enriched visualization. In their paper they presented a system, where for each word a vector is computed that contains the similarity values the word has with every other word. These vectors are used to build a high-dimensional similarity matrix, which allows the words to be laid out in the plane using multi-dimensional scaling (MDS). Their technique motivated further work into efficient systems to create semantic word cloud visualizations. While there are several algorithms to generate such word clouds with varied layouts~\cite{BarthKP14}, there is currently no interactive tool that allows the user to fine-tune these semantic word clouds in a human-in-the-loop way. 
	The online tool implemented by Barth et al.~\cite{BarthKP14} lets a user generate layouts using many different algorithms and allows for some limited dragging operations on the words, as well as deletion but neither operation is supported by algorithms that update the layout. Since the number of relevant word pairs in a typical data set exceeds by far the number of word pairs that can be adjacently placed by any planar word neighborhood structure, fully automated solutions must favor some of the many semantic links to be represented at the cost of missing others, purely based on numeric similarity scores. But only a human user, having deeper semantic knowledge about the underlying text, knows which of the chosen pairs are most relevant, which ones are missed, and which ones are less important. Hence, human-in-the-loop fine tuning can lead to more user satisfaction and semantic accuracy of the word clouds as meaningful text summary visualizations.
	While a tool like EdWordle~\cite{WangCBZDCS18} allows the user to easily edit a word cloud, having full manual control can be overwhelming for humans, and one should rather aim to combine the computational power of automated algorithmic solutions with the expert knowledge of the user.
	While the authors of EdWordle argue that a semantic word cloud could be used as input to their system to be edited, there is currently no simple way of importing a semantic word cloud into their tool. 
	Additionally, any updates on the layout are focused on preserving existing neighborhoods and the compactness of the layout. Since the algorithm has no knowledge of the semantic relationships of the individual words in the word cloud, a single displaced word could completely perturb the existing neighborhoods, leading to a flawed layout, where the user has no guarantee or feedback about its semantic quality.
	
	To address this gap, we introduce MySemCloud, an interactive human-in-the-loop semantic word cloud editor. Our tool generates a semantic layout of high quality using an algorithm inspired by Cui et al.~\cite{CuiWLWZQ10}, that performs well on the relevant evaluation criteria for word clouds~\cite{BarthKP14}. The main contributions of MySemCloud concern the subsequent editing steps and are two-fold. Firstly, we propose \emph{metric guides} which allow the user to visualize the underlying semantic relationships within the cloud, e.g., via the links between related words as shown in Figure~\ref{fig:teaser}, as well as guide the user towards potential improvements of the layout. Secondly, we propose \emph{semantic-enhanced} interactions. While most interactive systems focus on changing the appearance of the words, our system is focused on preserving the semantic quality of the word cloud. Accordingly, MySemCloud includes a smart dragging tool that considers the neighbors of dragged words and updates their position while preserving the visual stability of the layout. Additionally, extending EdWordle's compactifying layout updates, our system is able to consider the semantic relationships when updating the layout, which allows us to focus on maintaining the most meaningful adjacencies.
	
	In this paper, we first present an overview of the word cloud literature and current state-of-the-art layout algorithms (Section~\ref{sec:related_works}). Then, in Section~\ref{sec:semWC} we describe our layout algorithms, as well as the quality metrics we are designing our system for. In Section~\ref{sec:mscloud} we introduce the semantic-aware display as well a our semantic-enhanced interactions. Lastly, in Section~\ref{sec:eval}, we present an user study to evaluate our system. 
	MySemCloud can be accessed on \url{ac.tuwien.ac.at/mysemcloud}, where the code has additionally been made available.
	
	\section{Related work}\label{sec:related_works}
	
	Word clouds were introduced in the early 2000's~\cite{ViegasW08}, but were known then as tag clouds. In tag clouds, the words that occur the most in an input text are scaled proportionally to the frequency at which they appear in the data set and displayed in lines, laid out alphabetically or by descending frequency. To improve on these layouts, more compact display methods were proposed~\cite{SeifertKKGG08} that focused on packing the words more tightly. These compact tag clouds were quite similar to visualizations created by designers by hands, as for example \emph{pile of words} graphics that appeared in 2008 in the Boston Globe~\cite{schwenkler_2008}. Word clouds were further popularized by the Wordles website~\cite{ViegasWF09}, that allowed users to generate their own tag clouds, where the words are colored, displayed in a compact setting and sometimes rotated vertically. Appealing word clouds are often more than just compact layouts. For example, in Shapewordles~\cite{WangLCZBLZFHD20} the user chooses a (potentially complex) shape to draw the word cloud in. Extensions to the traditional word cloud design also include maps, where words are displayed on geographically significant areas~\cite{BuchinCLSW16,BhoreG0NW21,LiDY18}.
	
	\subsection{Semantic Word Clouds}
	
	The concept of using word placement in the plane is a logical way to encode more information within a word cloud. With semantic word clouds, the spatial distance between two words carries semantic meaning, namely placing closely related words nearby each other. While there exist different methods to compute a word similarity matrix~\cite{abs-1708-03569, CuiWLWZQ10, WuPWLM11}, our focus lies in the layout algorithms. Cui et al.~\cite{CuiWLWZQ10} proposed one of the first methods to generate such a semantic layout. Using multi-dimensional scaling (MDS), they computed 2D-coordinates for the words that approximate the desired relative distances to the other words. This technique usually results in a sparse layout. To avoid white space, they construct a Delaunay triangulation of the layout, on which they use a system of attractive forces preserving the neighborhoods. A similar result can also be achieved using a system of forces not on a triangulation of the layout but on the similarity graph itself~\cite{XuTL16}. Wu et al.~\cite{WuPWLM11} proposed an alternative to the force-based compaction by using seam carving. They identify vertical or horizontal sections of the drawing that are empty and remove them from the drawing.
	The semantic word cloud layout problem has also inspired research with a more theoretical focus. When representing the words by their rectangular bounding boxes, it is possible to transform the semantic word cloud problem to one reminiscent of rectangle contact graphs. In contact graphs, the edges of an underlying graph are meant to be realized by a proper edge contact between two boxes representing their respective vertices. In the Contact Representation of Word Networks (CROWN) problem, the number of edges of the input similarity graph that are realized in the contact graph has to be maximized. Barth et al.~\cite{BarthFKLNOPSUW14} proposed approximation algorithms to solve the problem on restricted graph classes which were later improved by Bekos et al.~\cite{BekosDFKKPSW17}. Barth et al.~\cite{BarthKP14} compared multiple semantic word cloud algorithms using the most common semantic word cloud evaluation metrics. To our knowledge, there is currently no semantic word cloud layout system that offers an interactive component. This is a natural extension of the model when considering the amount of interest that interactive word clouds have generated~\cite{JoLS15, KohLKS10, WangCBZDCS18}.
	The dynamic semantic word clouds introduced by Cui et al.~\cite{CuiWLWZQ10} have been studied further. They explored word clouds generated from a collection of documents at different timepoints. Not only was semantic relatedness encoded with proximity, but placement was also used to accommodate later changes of the data, e.g., words needing space to grow between two timestamps. Binucci et al.~\cite{BinucciDS16} designed a layout algorithm that creates such word clouds over time without a-priori knowledge of the complete data collection.
	
	\subsection{Interactive Word Clouds}
	
	Visualization construction and authoring tools are of great interest to the information visualization community. Thus, creating interactive tools was a natural next step to the growing popularity of static word cloud layout systems. One of the first tools created was ManiWordle~\cite{KohLKS10} which allowed the user to change the font, the color and orientation of the words as well as their position to potentially modify the whole layout. The interactive system allows users to fine-tune an automatically generated layout to match their personal aesthetic criteria. To maintain a compact layout, ManiWordle recomputes a layout using the Wordle algorithm but only considering the unedited words. In WordlePlus, Jo et al.~\cite{JoLS15} extended these interactive environments to multitouch systems. To update the word cloud after the changes, boundary words are moved in the gaps left by a potential edit. While the interactive component is positively received, updating the word cloud itself remains a challenge as both solutions tend to disturb the mental map of the user significantly. Wang et al.~\cite{WangCBZDCS18} proposed EdWordle, a solution using rigid body dynamics to preserve the neighborhoods of the unedited word as well as to preserve stability of the layout. While the authors of EdWordle argue that their tool allows users to edit semantic word clouds without destroying the layout, there is currently no straightforward method of generating an initial semantic layout with EdWordle, and additionally the user has no ability to conserve the semantic quality during the edits. If a user moves a word, its direct neighborhood is lost, and the user has no knowledge of whether or not significant semantic information was lost. Similarly, while the remaining neighborhood is preserved, there is no guarantee of its actual semantic quality. The visualizations that were created by designers using EdWordle are reminiscent of the semantic word cloud thematic clustering, which highlights their strength as an information communication tool, but to create those word clouds from scratch is time consuming and can be overwhelming. Also, if the user does not have expert knowledge of the text, they might not have the information necessary to create such a layout. Thus there is a need for a tool which provides a good quality semantic layout as a starting point that a user can easily fine tune subsequently, as well as information about the semantic similarities of the words in the text and interactions tailored towards preserving the semantic relationships in the layout.
	
	\section{Semantic Word Cloud layout}\label{sec:semWC}
	
	\begin{figure*}
		\centering
		\includegraphics{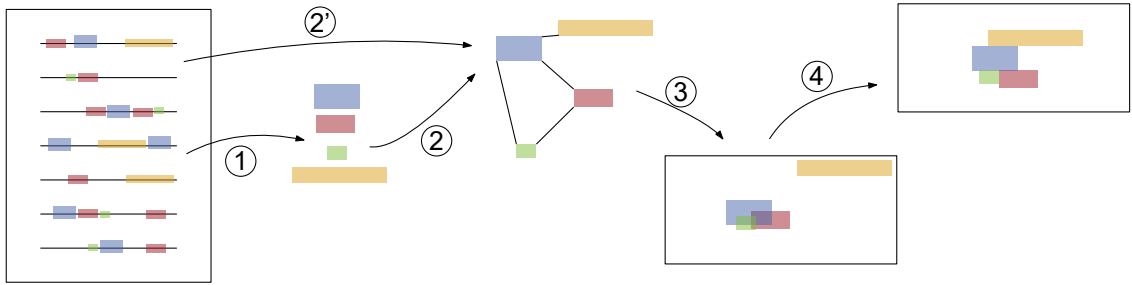}
		\caption{Steps to create an initial semantic word cloud layout: (1) stop words are removed, the remaining words are stemmed, and the top $k$ words are selected, (2) from those words and (2') the similarities computed from the co-occurrences in the input text
			(3) the words are laid out in the plane using MDS, and (4) attractive and repulsive forces are activated to obtain a compact, but overlap-free initial semantic word cloud layout.
		}
		\label{fig:pipeline}
	\end{figure*}
	
	The general problem of generating semantic word clouds has been studied in depth by Barth et al.~\cite{BarthKP14}. Here we present our method in detail, which has been inspired by several works and tailored to our needs and implementation choices (JavaScript and D3.js).
	There are two main steps to the generation of a semantic word cloud layout. The first step involves generating from an input text a similarity graph and the second step concerns the actual layout of the words in the plane. An overview of our system can be seen in Figure~\ref{fig:pipeline}. 
	The goal is to generate a layout, where the relative position of two words indicates their similarity or lack thereof. There exist multiple metrics to describe the relationships between the words in the input text, and in the visualization, word pairs that score highly on the chosen relatedness metric should appear closer together than word pairs with a low score. This results in thematic clusters in the final layout, where the user is then able to understand the content and how different terms are related in the text source.
	
	\subsection{Constructing the semantic similarity graph}
	
	The first step is to extract the words from the given text, which will later form the vertex set of our similarity graph. Using the natural language processing (NLP) library Natural~\cite{umbel_ellis_mull_2011}, we first remove irrelevant words from the text, for example ``that'', ``the'', ``for'', ... that are not significant and should not be displayed. We then use stemming on the remaining words, meaning words are shortened to their meaningful stem, e.g., the words ``explanation'' and ``explaining'' would both be understood by the algorithm as their stem ``expla''. For each of these stems we choose one of the words of the text as the representative of all the words with the same stem. This allows us then to accurately rank word frequencies, and to choose the top $k$ words that are found (see step (1) in Figure~\ref{fig:pipeline}). We found that in most cases $k=50$ is sufficient to cover the main themes of the text without overloading the visualization. This step is already sufficient to create a word cloud without additional semantic information. Finally for each chosen word, the number of times the stem occurs is summed up and we scale the font size of the word proportionally to this frequency.
	
	To evaluate word similarities, further pre-processing is required. For a pair of words $w_1, w_2$, their similarity is calculated using the Jaccard similarity, which performs slightly better than the cosine similarity~\cite{BarthKP14}, a common alternative. 
	We calculate it in the following way. Let $S(w)$ be the set of sentences the word $w$ appears in. Then the Jaccard similarity of two words $w_1,w_2$ is a score in $[0,1]$ given by:
	\begin{align*}
		s(w_1,w_2)=\frac{|S(w_1)\cap S(w_2)|}{|S(w_1)\cup S(w_2)|}.
	\end{align*}
	From this we create a complete weighted graph $G=(V,E)$, where the vertices in $V$ represent the top $k$ words selected and the edges are weighted by the Jaccard similarity $s(w_1,w_2)$ for each edge $(w_1,w_2) \in E$. Some word pairs might have little or no similarity, meaning the two corresponding words rarely occur in the same sentence; thus we remove edges with Jaccard similarity below some threshold $\sigma$ from the graph. We set $\sigma=0$ as a default, removing only edges corresponding to words that never co-occur (see step (2) in Figure~\ref{fig:pipeline}).
	
	\subsection{Creating the initial layout}\label{subsec:layout}
	
\begin{figure}
		\centering
		\includegraphics{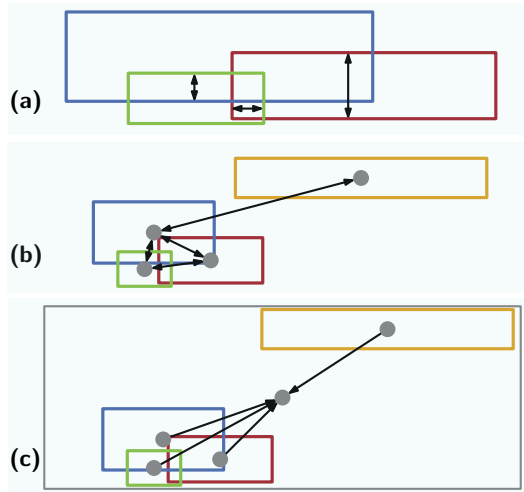}
		\caption{The force system to create the initial layout, \textbf{\textsf{(a)}} shows the repulsive forces that push overlapping words away from one another to prevent the bounding boxes from overlapping, \textbf{\textsf{(b)}} the attractive forces that pull all words toward their similar neighbors, proportionally to their respective similarity score and \textbf{\textsf{(c)}} the attractive forces that pull all words towards the center, 
		}
		\label{fig:force}
	\end{figure}

	Using multidimensional scaling (MDS)~\cite{GansnerKN04}, we initialize the positions of the vertices of $G$ in the plane. At this step some words may be tightly clustered together and overlap, while others may be spread much further apart (see the example layout of step (3) in Figure~\ref{fig:pipeline}). We then apply a forced-based system to the graph to adjust word positions, meaning we assign forces to the edges and vertices of the graph and then use these forces to simulate the motion of the vertices. It is important that words do not overlap one another so we must consider node overlap removal methods~\cite{ChenPPS20} that rely on a force system to naturally extend our layout algorithm. We calculate the distance between each pair of words in the x and the y dimension (see  Figure~\ref{fig:force}a). If their bounding boxes overlap, the words are either pushed vertically or horizontally away from one another, in the direction that resolves the overlap fastest. That is, if there is more overlap in the x-dimension, the force will be applied vertically, thus ensuring a top/bottom or side contact between the two boxes remains after the resolution of the overlap (see step (4) in Figure~\ref{fig:pipeline}). This method is similar to the Force-Transfer-Algorithm introduced by Huang et al.~\cite{HuangLSG07}. 
	
	To obtain a compact layout, we must also apply attractive forces between each pair of words that is connected by an edge in $G$ as well as a centering force on every word. Consider a word $w_1$ that shares an edge with a word $w_2$. Then there is a force from the center point of the bounding box of $w_1$ oriented towards the center point of $w_2$ that is scaled by the value $s(w_1,w_2)$, meaning more strongly related words have a stronger attractive force (see  Figure~\ref{fig:force}b). To ensure that our layout is displayed in the center of the canvas, we also add a weak attractive force from each word to the center of the canvas (see Figure~\ref{fig:force}c). Since our graph is dense, the force system might struggle to find a stable layout immediately, thus, we let it recompute iteratively new positions while decreasing the strength of the forces before we stop its computation and we obtain a final layout. Since our system is interactive, it is undesirable to let the system stabilise itself for too long as it affects the responsiveness of our system negatively. But it is also necessary to not stop it too early either to ensure our layout is of sufficiently good quality. We found experimentally that 1000 iterations, which could be computed in about 230ms, are a good compromise.
	
	\subsection{Semantic word cloud quality metrics}\label{subsec:metrics}
	
	To design an automatic word cloud system, we often rely on optimizing quantitative metrics that measure the aesthetic qualities that are desirable in such a visualization. For classical, non-semantic word clouds, creating a compact design is the main visual criterion, and it remains an important aspect for semantic word clouds, too. But we must also consider metrics for the semantic quality of the visualization. Next, we introduce the main quality metrics that are relevant to evaluate semantic word cloud layouts~\cite{BarthFKLNOPSUW14,BarthKP14,BekosDFKKPSW17}.
	
\begin{figure}
		\centering
		\includegraphics[width=0.8\linewidth]{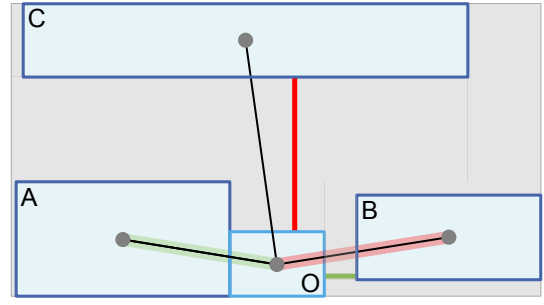}
		\caption{Word O is strongly connected to every other word. It successfully realizes its adjacency with word A (green highlight), but not with the word B (red highlight) as the boxes do not touch, but the distortion value with word B is good (green edge) as they are still close, unlike the distortion value with word C (red edge). The compactness corresponds to the ratio of the blue area over the area of the gray bounding box.}
		\label{fig:metric}
	\end{figure}

	The semantic quality can be measured in two main ways, the first being \emph{realized adjacencies}~\cite{BarthFKLNOPSUW14,BarthKP14,WangCBZDCS18}. When modeling the words as their rectangular bounding boxes, a realized adjacency corresponds to a segment contact between two boxes that share an edge in the semantic graph, see the highlighted edges in Figure~\ref{fig:metric}. As this edge is weighted, the metric is weighted as well, meaning that it is better to realize a contact between two highly related words over one or more contacts of  low weight. This metric captures the notion that the simplest way to understand that two words are related is if they are directly next to one another. As the boxes themselves are an abstraction of the words, rather than checking for a proper contact between two boxes, we check if the bounding boxes of two words $w_1$ and $w_2$ overlap. More precisely, assume $w_2$ has the smaller bounding box. We artificially inflate the size of its bounding box by 20\% and check if it overlaps with the bounding box of $w_1$. Experimentally, we found that limiting a contact to a distance of 0 between two bounding boxes was too strict and words that visually appeared to be in contact were not counted as not realized.
	For a semantic input graph $G=(V,E)$ and a word cloud layout $\Gamma$ of $G$, the value $r(\Gamma) \in [0,1]$ of the realized adjacencies $E' \subseteq E$ according to the above definition is calculated in the following way:
	\begin{align*}
		r(\Gamma)=\frac{\sum_{e\in E'}s(e)}{\sum_{e\in E}s(e)},
	\end{align*} 
	where $s(e)$ is the similarity score (weight) of the edge $e$ in $G$. To realize every adjacency, the input graph would need to be planar~\cite{BuchsbaumGPV08}. But semantic similarity graphs are dense, almost complete graphs, so this value $r(\Gamma)$ is often low. Nevertheless it is still an effective method to compare two layouts as that value can easily double from one drawing to the other. For an arbitrary graph $G$, finding the maximum realizable adjacency value is an \NP-hard problem~\cite{BarthFKLNOPSUW14}. It was found that the cycle cover algorithm~\cite{BekosDFKKPSW17} has the best performance for this metric~\cite{BarthKP14}.
	
	The second semantic quality metric is \emph{distortion}~\cite{BarthKP14}, which compares the distance of each word pair to their similarity score. Distortion can be seen as a relaxation of realized adjacencies as two highly correlated words do not need to touch but can instead be sufficiently close to indicate a meaningful relationship in the visualization, as shown by the colored edges in Figure~\ref{fig:metric}. It also reflects the notion that unrelated words should not be close to one another, which the realized adjacencies metric fails to properly account for, since in its commonly accepted definition there is no penalty when two unrelated words touch. But with distortion, if their similarity value is low then they should be far away from one another in the plane. The distortion value $d(\Gamma)$ of a layout $\Gamma$ of $G$ is computed using Pearson's correlation coefficient $\delta(\Gamma)$ between the (dis)similarity matrix and the distances realized in the plane: 
	\begin{align*}
		\delta(\Gamma)=1-\frac{\sum_{(u,v)\in E}(1-s(u,v)-\overline{(1-s)})(d_{\Gamma}(u,v)-\overline{d_\Gamma})}{\sqrt{\sum_{(u,v)\in E}(1-s(u,v)-\overline{(1-s)})^2(d_{\Gamma}(u,v)-\overline{d_\Gamma})^2}},
	\end{align*}
	where $1-s((u,v))$ corresponds to the dissimilarity of $u$ and $v$, $\overline{(1-s)}$ is the average dissimilarity value in $G$ and similarly, $d_{\Gamma}(u,v)$ is the minimum distance between the bounding boxes of $u,v$ in $\Gamma$ and $\overline{d_\Gamma}$ is the average distance in $\Gamma$. The distortion is then defined as
	\begin{align*}
		d(\Gamma)=\frac{\delta(\Gamma)+1}{2},
	\end{align*}
	as $\delta(\Gamma)$ has its values in $[-1,1]$. A value of $d(\Gamma)=1$ indicates that every word is positioned at an ideal distance from any other word, $0$ indicates the inverse and a value of 0.5 signals that there is no correlation between similarities and distances. 
	Barth et al.~\cite{BarthKP14}, who first introduced the distortion metric, found that the seam-carving algorithm~\cite{WuPWLM11} performed well with this metric.
	
	Both metrics help us gauge the semantic quality of a layout. To evaluate its overall aesthetic quality, one can further consider \emph{compactness}~\cite{BarthKP14,WangCBZDCS18}. The compactness $c(\Gamma) \in [0,1]$ of a layout $\Gamma$ of $G$ represents the ratio of the space used by the words over the total space available in the bounding box of $\Gamma$. More precisely, 
	\begin{align*}
		c(\Gamma)=\frac{\sum_{v\in V}a(v)}{a(\Gamma)},
	\end{align*}
	where $a(v)$ represents the area of the bounding box of $v$ and $a(\Gamma)$ is the area of the  bounding box of the entire word cloud. 
	Most non-semantic word cloud layout methods achieve high values of compactness as they can form a tight packing without considering relative word placement. When semantics are considered, the need to separate some words from each other to obtain good distortion values often leads to lower compactness values.
	
	There exist further metrics~\cite{BarthKP14} to evaluate the visual quality of a word cloud layout, namely \emph{uniform area utilization}, which requires the words to be evenly distributed over the canvas. One can also consider the \emph{aspect ratio} of the layout, where a ratio of 1 could be desirable, or one closer to traditional media formats like 16:9, in contrast to extreme aspect ratios, which might make the visualization difficult to read. 
	
	As our main interest point is the semantic quality of the layout and how it can be maintained during interactive steps, we will focus our attention on the realized adjacencies and distortion values of our layout, but retain the compactness metric in our quantitative evaluation as it is the most established of the three aesthetic layout quality metrics.
	
	\section{MySemCloud}\label{sec:mscloud}
	
	MySemCloud is designed as a tool for a general public audience, with no deeper design expertise. It is created for users who wish to summarize familiar texts with informative word clouds in different media forms (presentation, social media) and focus on the information delivery. The target user is expected to have expert knowledge of the input text, but limited expertise in graphic design. MySemCloud should be simple to use to ensure the user can quickly achieve a desired result. Therefore we focused on repositioning operations that are directly made on the visualization canvas. It is meant to be an alternative to existing editing tools which focus heavily on aesthetics and extensive design expertise.
	
	In this section we present the technical details of our novel interactive semantic word cloud editing tool MySemCloud. Typically, a word cloud for a given text computed using the approach outlined in Section~\ref{sec:semWC} is of good quality but does not take subjective user preferences into account yet. %
	Since current algorithmic methods cannot predict which semantic relationships a user prefers to highlight over others, MySemCloud implements intelligent interaction modes and visual aids meant to guide the user in fine-tuning the computed word cloud themselves.

	To help the user during their desired editing steps, MySemCloud provides two smart support tools, a \emph{semantic-aware} display and \emph{semantic-enhanced} interactions. %
	
	\subsection{Semantic-aware display}
	
	\begin{figure*}
		\centering
		\includegraphics{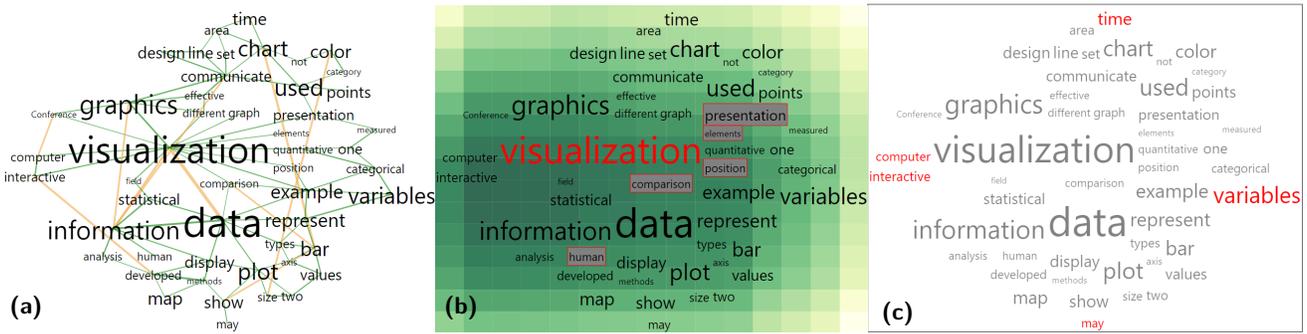}
		\caption{The different metric guides, \textbf{\textsf{(a)}} shows the realized adjacencies guide with the realized edges in green and the strongest missed adjacencies in yellow, \textbf{\textsf{(b)}} shows the heat map that indicates the positions with the highest distortion values for the selected word ``visualization" in red and highlights the five most misplaced words, and \textbf{\textsf{(c)}} shows the compactness metric guide with the words stretching the bounding box highlighted.
		}
		\label{fig:mguides}
	\end{figure*}
	
	The \emph{metric guides} are a system of display layers meant to translate the underlying semantic data into visual cues that help the user understand and edit the layout while optimizing its visual and semantic quality. Specifically they correspond to three options that each can be toggled on or off together or separately and change the user interface. %
	
	\paragraph{The adjacencies metric guide} lets the user display the edges of the semantic word similarity graph. When toggled on, for every edge in the semantic graph whose endpoints are reasonably close to each other in the visualization to be considered an adjacency (see Section~\ref{subsec:metrics}), the edges are displayed using a green segment between the center points of the two words corresponding to the edge's endpoints. The width of the segment is scaled proportionally to the similarity value of the edge. Additionally, some missing adjacencies are also shown in yellow. Since the graph is very dense, we only focus on showing the most significant missing contacts in the graph. We sort the list of missing adjacencies and select the ten highest-weight edges that are not realized, as shown in Figure~\ref{fig:mguides}a. This helps the user see where strongly connected components lie as well as the main missing adjacencies. %
	
	\begin{figure}
		\centering
		\includegraphics[width=0.6\linewidth]{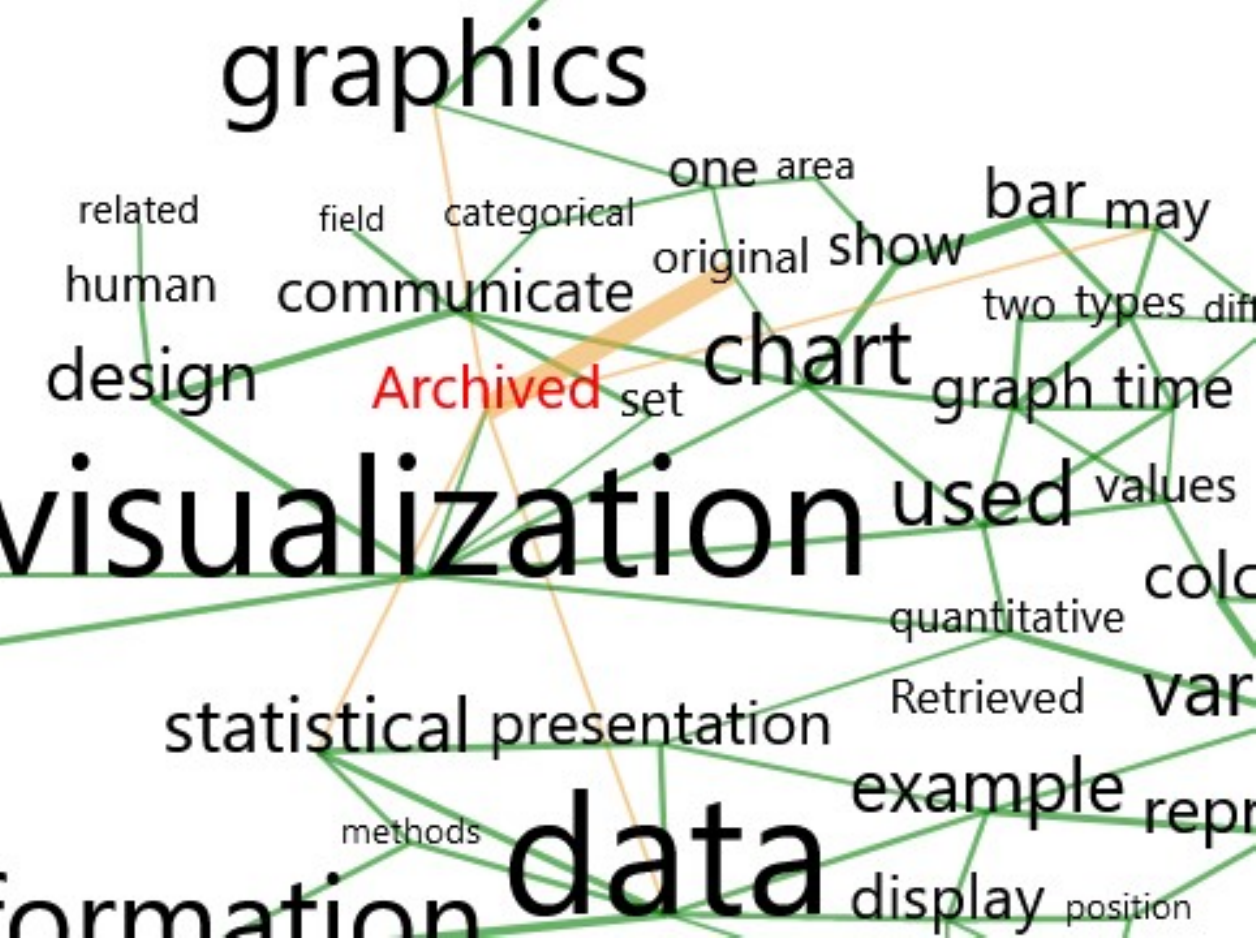}
		\caption{The right click operation under the adjacencies metric guide show the five links of the selected word ``Archived''. We see one realized adjacency with ``visualization'' and five missing ones.
		}
		\label{fig:adjguide}
	\end{figure}

	To ensure that all the data is visible, we offer a secondary view: when selecting a specific word using the right click, we replace the display of the main missing adjacencies of the cloud, with a display of all the edges that are incident to the selected word %
	as shown in Figure~\ref{fig:adjguide}.
	When the adjacency metric guide is active, if a user wants to move a word, they have the information of the word's adjacencies in its starting position, adjacencies that will likely be lost, and the adjacencies that might be realized in its new neighborhood, and thus make a decision of how to best adjust the position of a word. The user might also decide which missing adjacencies are globally too important not to be realized and quickly identify those in the general view.
	
	\paragraph{The distortion metric guide} shows to the user which positions are, or are not, semantically meaningful. When selecting a word with the right click in the distortion view, a heat map will be displayed, where a darker green shade indicates positions that achieve high values for the distortion metric for the chosen word, and pale yellow shades represent positions that realize low values for the same metric.
	To create the heat map, we produce a tiling of the bounding box that the word cloud currently occupies. We compute a new distortion value by updating the length of the edges incident to the selected vertex only, then use a color gradient to associate a shade with the values obtained on a scale from dark green to pale yellow. This creates a color scale that indicates in which areas of the cloud the strongest neighbors of the selected word lie, and in which area the unrelated words are. The resulting view can be seen in Figure~\ref{fig:mguides}b.
	When turning the metric guide on, the words with the most negative impact on the distortion are highlighted in grey. To compute this, we calculate for each edge $e$ of the input graph an \emph{ideal length} $\ell(e)=(1-s(e))\frac{D}{2}$, where $D$ is the longest distance between two words in the visualization. 
	We calculate the \emph{penalty}, i.e. the difference between the ideal length and the actual length of $e$. If two words are unrelated and the difference is positive, they are too close and the penalty is squared. For each word we sum the penalties incurred with all other words, and finally highlight in grey the five words that achieve the highest sum.
	
	\paragraph{Lastly, the compactness metric guide} helps the user create a more space-efficient layout. When active, the bounding box of the word cloud is displayed, and the words that are on the boundary are highlighted (see Figure~\ref{fig:mguides}c). A user interested in creating a more compact layout can then select a boundary word and, using any of the two semantic metric guides, find a new suitable position that results in a more compact layout. This guide can be used alone, but as it only optimizes towards compactness, it is more relevant for semantic word clouds when used alongside the distortion or adjacencies guide. When used with a semantic guide, the user can more easily consider the global appearance of the word cloud while improving its semantic quality. %
	
	\medskip
	The three different views can be used individually or in any combination. Thus, the user can choose how to edit the layout in a way that can preserve the important neighborhoods, or they can choose to improve it by using the information from the underlying data set. While some views could potentially contain more information, we chose to prioritize simpler views to encourage the user to layer views on top of one another.

	\subsection{Semantic-enhanced interaction}
	
	The default interaction step of MySemCloud is a drag-and-drop operation, where the user moves a word from one place to another. Once the new word is in place, we resolve any overlaps  that were induced by the move. These edits cause minimal disturbances of the user's mental map and are useful for precise refinements of the layout.
	We additionally provide two semantic-enhanced edit modes. Updating a force-based layout is often a challenge: a single local move can greatly perturb a graph's layout. Thus, our semantic-enhanced interactions are not only focused on incorporating the semantic graph with the interaction, but also on maintaining the stability of the previous layout.
	After any move is done by the user, the values of the quality metrics of the layout are updated to give the user direct feedback about how much impact the move they made has had.
	
	\paragraph{The \emph{neighborhood-follows} mode} helps preserve the distortion and adjacencies in the graph. Specifically, when this mode is toggled on and a word is moved, the positions of its strong neighbors will be updated as well. There are two main components to this step.
	
	The first component concerns the selection of the vertices whose position should be updated. We compute a breadth-first search tree of $G$ rooted at the moved word $w_1$, and add vertices from that tree to the \emph{relevant vertices list} in the following way. 
	We first consider all the children of $w_1$. If those have similarity value with $w_1$ higher than our threshold $\theta$, they are added to the list. 
	We then look at the children of those selected vertices. We compute the ratio of their similarity value with their parent over their depth in the tree. If those are higher than $\theta$, they are similarly added. We continue down the tree using the same ratio of similarity with the parent divided by depth of the vertex until all vertices have been considered.
	This method ensures that we are less likely to select grandchildren of $w_1$, and will do so only if they are strongly related to their parent (and that parent is strongly related to $w_1$). We set $\theta=0.1$.

	The second component is the update of the layout itself. First the non-overlap forces and attractive forces are turned off. An \emph{anchoring} force is added to every word, oriented from the center point of the word to its current position at the start of the move. For each edge linking a vertex of the relevant vertices list to their parent, we reactivate the attractive forces in the following way: all edges between $w_1$ and its children in the list are reactivated with their full strength, all the edges between the chosen children of $w_1$ and their own chosen children will be reactivated at partial strength, and every further edge will have its strength decreased proportionally to their depth in the BFS tree of  $w_1$.
	
	\begin{figure}
		\centering
		\includegraphics[width=0.8\linewidth]{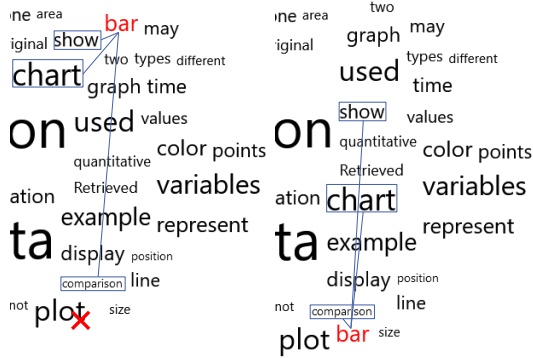}
		\caption{``Bar'' is moved to the position marked by a red cross using the neighbors follow mode, the algorithm selects ``show", ``chart" and ``comparison" as its significant neighbors. ``Comparison" was close to the new position and is able to realize the contact after the move. ``Chart" and ``show" also move closer to ``bar" but ``show" has more similar neighbors in the upper part of the visualization and thus does not move too far away.
		}
		\label{fig:nfollow}
	\end{figure}
	
	Finally, when the new positions are computed, the anchoring forces of the moved neighbors are updated to be directed to their current position, the overlap removal forces are reactivated and we compute the final layout. Using this move, when a word is dragged, its highly similar neighbors will follow it, thus preserving the important adjacencies in the graph. Since those following words might have strong adjacencies themselves, we search deeper in the tree to find if some are significant enough that further words might be moved as well. The anchoring force pulls any following words back towards their starting position. This helps to maintain the stability by not permitting the moved words to go too far. It also avoids significantly disturbing the distortion value of the layout as can be seen in Figure~\ref{fig:nfollow}.
	
	\paragraph{The fill-holes mode} aims to preserve the compactness without damaging adjacencies. A common issue with interactive word clouds is that when a word is moved, the space left behind should be filled to re-establish the compactness of the layout. In MySemCloud, we resolve this issue by reactivating the forces of the system similarly to how we generated the first layout. Specifically all the attractive forces corresponding to the edges are reactivated, as are the centering and the non-overlap forces. This allows the layout to re-compactify itself, in a manner where words that are more strongly related to one another will more likely be pulled together into the hole created by a word move than weakly related words. This operation can also be triggered with a button, without needing a move.
	
	\medskip

	When used together, these interactive modes tend to update the layout significantly such that a user's mental image might be disturbed. The fill-holes mode %
	tends to increase the value of realized adjacencies, as it attempts to close the gaps between words. %
	The neighbors-follow mode can also be used for larger updates%
	: when wanting to place a new topic in an entirely new area of the layout, it can effectively allow the user to move a cluster of related words at once.
	
	MySemCloud further contains a mode to toggle the bounding boxes of the words, as it makes it easier to notice directly if the contacts are realized or not. This is also helpful with compactness as one can quickly spot gaps in the rectangle packing. 
	It also contains an undo button which reverts the layout to its state before the last move was executed, as well as a button to save a certain state of the layout. The user can then load any saved state at any point in the editing process, or recover an unsaved state using the undo button. Lastly, the values of the metrics are displayed for the current layout, the previous layout, as well as the best value achieved by a layout.
	After every move the values are updated and if the layout achieves a new optimum for any value, it is saved automatically. %

	\section{Evaluation}\label{sec:eval}
	
	We evaluate MySemCloud from two different perspectives. First, we explore in a controlled study how the semantic-aware display and the semantic-enhanced interactions are able to allow users to improve the quality of initial semantic layouts. Second, we present findings from a qualitative study during which participants were able to freely use MySemCloud to design word clouds of their own text data.
	We want to show that MySemCloud is not only able to generate word clouds of high semantic quality, but that it additionally is a good compromise between the existing interactive, but non-semantic word cloud editors and the non-interactive, semantic word cloud layout algorithms.
	
	We implemented MySemCloud in JavaScript. The text submitted in the client is sent to a backend server running on Node.js that handles the semantic similarity computation using the NLP library \textit{Natural}~\cite{umbel_ellis_mull_2011} and generates the MDS layout. 
	The final layout is computed in the client using the popular JavaScript visualization library \textit{D3.js}~\cite{bostock} for the force layout computation and the rendering. %
	
	Both aspects of the evaluation were performed as a back-to-back in-person user study that took 45min during which each participant worked individually with the tutor. We recruited 20 participants (5 women, 15 men) who were students or researchers in Computer Science. None of the participants had used a word cloud creation tool or layout algorithm previously. One participant reported not being familiar with word clouds, and another reported having heard about semantic word clouds. All participants reported normal or corrected-to-normal vision, and had no color vision deficiencies. The study was conducted on a 27" LCD screen at a $2560\times1440$ resolution using a mouse as input device.
	
	\subsection{High quality layout creation}
	
	We introduced in Section~\ref{subsec:metrics} several metrics to evaluate the quality of a semantic word cloud layout. In the study, we focused on the two semantic quality metrics \emph{realized adjacencies} and \emph{distortion}, and on the non-semantic \emph{compactness} metric. In this section we evaluate the quality of the layouts created by the 20 participants using MySemCloud. Specifically, we investigate how efficiently users can improve the quality of the layouts using the  semantic-aware display and semantic-enhanced interactions provided in MySemCloud.

	\subsubsection{Study Design}
	
	The study was conducted in three steps, an introduction and training phase, followed by a set of word cloud improvement tasks, and completed by a short questionnaire.
	
	To start, the participants were given an introduction to semantic word clouds, as well as the definition and intuition behind each metric. They were then introduced to the tool, and each of its functionalities was explained.
	They were given time to learn to use the tool, and told what the tasks would consist of. During this training they could ask questions and familiarize themselves with MySemCloud.
	
	\paragraph{Data Sets.} The data sets used for the study were the following: (1) A summary of the book ``Harry Potter and the Philosopher's Stone''~\cite{mohanda} as a training data set and (2) the English Wikipedia page for the ``European Union''~\cite{wikipedia_2022} for the study tasks.
	
	As the users of MySemCloud are expected to have some familiarity with the texts they are designing a cloud for, we chose text data sets covering topics of broader public interest.
	One participant reported no familiarity with the Harry Potter book, but all participants reported having sufficient knowledge of the European Union to understand the content of the layout being presented to them.
	
	\paragraph{Tasks.} Four tasks were given to the participants to evaluate the different aspects of our tool. They were asked to improve the value of a metric as much as possible within a given time. For task 1 they had to improve the \emph{realized adjacency} metric,  for task 2 the \emph{distortion} metric, the \emph{compactness} metric for task 3 and to improve in parallel as much as possible the values of the \emph{realized adjacency} and \emph{compactness} metrics for task 4.
	While the participants spent time training on the tool before the tasks were undertaken, we randomized tasks 1 through 3 to avoid systematic bias through leaning effects. We gave the participants 2 minutes for tasks 1, 2, and 3 and 2.5 minutes for task 4.
	We calculated the improvement rate achieved by each participant within each task for the targeted metric. We did not give participants a lot of time as we were interested at how efficient our design was at guiding the users during the edits. The last task was given more time as we expected it would be more complicated for the participants to optimize two potentially conflicting goals simultaneously. We focused on combining compactness and realized adjacencies as those have been the strongest focus of previous semantic word cloud layout algorithms~\cite{KohLKS10,WangCBZDCS18,BarthKP14,BarthFKLNOPSUW14,BekosDFKKPSW17}.
	The participants were not obliged to use the relevant metric views for each task or specific interaction modes, but were asked to choose the setup that they felt most efficient working with.
	
	Once the participants had completed the tasks, they were given a questionnaire to describe their understanding of the quality metrics and to evaluate the difficulty of the task.
	
	\subsubsection{Results/Findings}
	
\begin{figure*}
		\centering
		\includegraphics{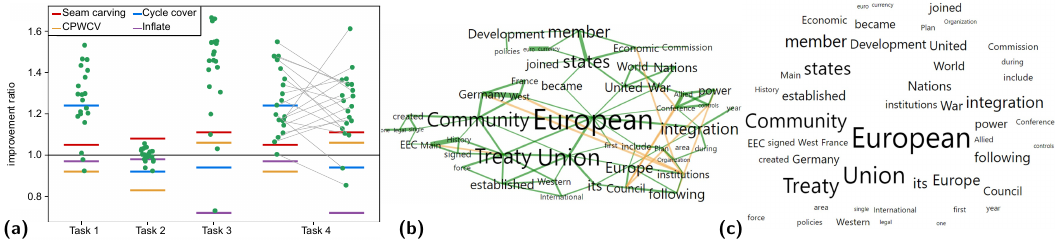}
		\caption{\textbf{\textsf{(a)}} Each point corresponds to the ratio of the metric value obtained by a participant for the task indicated by the column over the value of the starting layout. The left values in task 4 correspond to realized adjacencies and on the right is the compactness. The grey lines link values obtained in the same layout. A value of 1 means no improvement, anything above the line corresponds to a better value in the targeted metric and a value below 1 shows a worsening of the score.
			On average every task was successful, although distortion was the most difficult overall. The horizontal lines correspond to values achieved with the algorithms specified in the legend~\cite{BarthFKLNOPSUW14,KohLKS10,BarthKP14,WangCBZDCS18}. \textbf{\textsf{(b)}} an example of a semantic word cloud achieving high realized adjacencies values and an improvement ratio of 1.84, \textbf{\textsf{(c)}} an example of a semantic word cloud achieving high distortion value corresponding to an improvement ratio of 1.11. Both layouts were created using MySemCloud in under 5 minutes and outperform the best automated layouts.
		}
		\label{fig:task1}
	\end{figure*}

	Figure~\ref{fig:task1}a shows the results of the four metric improvement tasks. Our hypothesis was that we believed candidates would successfully complete all four tasks, but struggle with task 4 when balancing the two different metrics. We found that the candidates were most successful with the compactness improvement task, as 19 participants improved the compactness of the bounding box. For the task of increasing the realized adjacencies values, the candidates were similarly very successful, with only two candidates failing to improve the layout. 
	When the candidates were tasked of improving adjacencies and compactness in parallel, they were similarly successful. No participant failed to improve realized adjacencies on this second attempt, but some neglected the bounding box improvement. 
	
	Three candidates had chosen for the compactness task to completely ignore any semantic positioning by creating a tight packing of rectangles, but could not rely on this strategy on the combined task. Those participants successfully improved the layout when they had to preserve and improve its semantic quality.
	
	The participants on average clearly outperformed the four automated layout algorithms on most tasks. The cycle cover algorithm, that is designed towards maximizing realized adjacencies was outperformed by 11 participants in task 1 as well as in task 4. No participant was able to obtain a better distortion value than what was achieved by the seam carving algorithm, but they usually performed better than most of the competing algorithms, confirming the anticipated advantages of our human-in-the-loop approach.
	
	We also note that they were able to efficiently target the two metrics at the same time in task 4, unlike for example cycle-cover which is targeted towards adjacencies and performs poorly on compactness. Candidates in general did not heavily disregard one metric for the other as the best participants achieved good values for both metrics and similarly the worst performers tended to struggle with both metrics.
	
	The task to improve distortion proved difficult, as eight participants did not succeed in improving the distortion value of the initial layout. During the interview, six participants reported that while they understood the intuition behind this metric, they had issues understanding how to translate it visually. When asked which metrics they thought were the most relevant for semantic word clouds, distortion was the best received metric with eight participants commenting that it was the most relevant metric for these layouts, two of those had reported having trouble with the distortion improvement task. 
	
	The difficulties with this task are likely due to the participants focusing on the larger words of the layout. Those, when moved, tended to heavily disrupt the initial layout, often lowering its overall quality. More successful participants focused instead on average and smaller sized words, which often were farther from ideal positions. Given additional targeted training and more time for the task, we suppose that candidates could have been more successful. An example of a word cloud with high realized adjacencies and distortion scores can be found in Figures~\ref{fig:task1}b and~\ref{fig:task1}c. Notice that high value distortion layouts are less densely packed, which is contrary to high realized adjacency and high compactness value layouts. As many participants naturally tried to augment the compactness of each layout even when the tasks did not require it, this can also explain the difficulties they encountered.
	One might consider an alternative definition to the metric, that is compatible with denser layouts, or that results in higher swings of the value of the metric which would give more feedback to the user regarding the efficiency of their edit.
	
	We hypothesized candidates would understand the compactness metric most easily, as well as not have many difficulties with the realized adjacencies metric. Each participants graded the difficulty of improving each metric on a seven point Likert scale, where 1 meant the task was very difficult and 7 meant very simple. They found that compactness was the simplest, rating it a 6.3 (simple), distortion was the most difficult giving it a 3.95 rating (neither simple nor difficult), and they gave realized adjacencies a 4.65 grade (slightly simple to simple).
	
	\subsection{Qualitative evaluation}
	
	In this section, we want to understand how MySemCloud could be used as a word cloud design tool and how users approach it as a visualization tool to present a text of their choosing. 
	We want to study the interest users have in semantic word cloud layouts and how much they value being able to edit and fine tune their layouts.

	\subsubsection{Study Design}
	
	For this task, we simulated  a typical use case of our tool, where a user creates a semantic word cloud for a text they have a deep knowledge of.
	As our pool of participants contained 13 researchers and 7 students who had at least obtained a Bachelor's degree, each participant was able to choose a scientific text that they had expert knowledge of. Of the 20 participants, 10 chose a paper they were a main author of, 4 chose their thesis, and 6 chose a paper they had thoroughly studied.
	
	A semantic word cloud was generated from the chosen input text using MySemCloud, and the participants were asked to edit the layout into a visualization of their liking.
	They were able to take as much time as they desired and could use any functionality of MySemCloud. Overall, the candidates took between 5 and 20 minutes to arrive at a final layout and had different design goals.
	
	The candidates were then introduced to the semantic word cloud generation tool created by Barth et al.~\cite{BarthKP14}. The functionalities were explained and they were able to try out the different algorithms on the text they had previously chosen. They were also shown the layouts generated using the Cycle Cover algorithm~\cite{BarthFKLNOPSUW14}, the layout computed using the Seam Carving method~\cite{WuPWLM11}, the Inflate \& Push layout~\cite{BarthKP14} as layouts which, respectively, achieved high values for the realized adjacency, distortion and compactness metrics.
	
	They then completed a questionnaire covering their experience using MySemCloud, their impression of semantic word clouds in general and they were asked to compare MySemCloud to the non-interactive layout algorithms. 
	Lastly, an interview was conducted during which each participant was asked about their design goals for their personal word cloud, their impression of the metrics and the quality of the visualization as well as their impression of MySemCloud.
	
	\subsubsection{Results/Findings}
	
	Our hypothesis was that participants would preserve the semantic grouping created by the original layout, and would focus on changing the placements of some words to more appropriate topic clusters.
	\paragraph{Design goals.}  We identified three different types of design goals amongst the participants: the compact designs (13), the clustered designs (5) and the mind map designs (2).
	
\begin{figure*}
		\centering
		\includegraphics{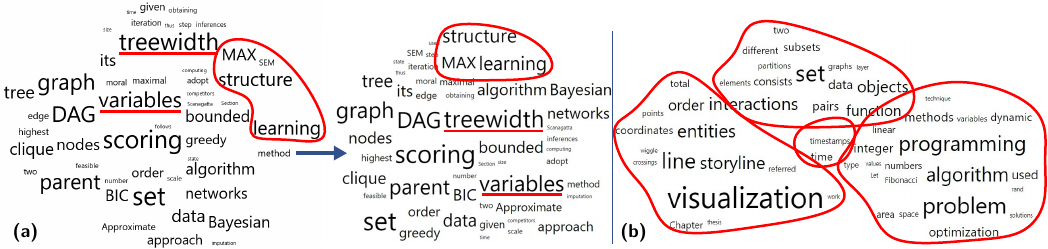}
		\caption{Example of semantic word clouds generated for task 2, \textbf{\textsf{(a)}}: a word cloud before and after the user fine-tuned the layout, the larger words have been organised in topic clusters and few smaller words have been moved, \textbf{\textsf{(d)}}: a word cloud created by a user, gaps separate the different themes which are themselves clusters together.
		}
		\label{fig:task2}
	\end{figure*}

	An example of a compact design can be seen in Figure~\ref{fig:task2}b. Here the participant did not edit the initial layout (Figure~\ref{fig:task2}a) significantly, most of the changes are results of the interaction modes and overlap removal forces. The main aspects of those compact designs revolve around a strategic placement of the largest words toward the center. Our algorithm tends to draw the bigger words towards the center as they often have a very high degree in the similarity graph. This was rated positively by eight participants as it aligned with their design goal. The smaller words are naturally arranged on the periphery. Some of those words are moved, often using the hints given by the semantic metric guides, closer to the most related large neighbor. In the interviews, eight participants describe a layout with the main themes centered as an ideal layout. The resulting layouts appeared more compact, but due to their often rounded designs achieved on average a coverage of under 60\% of the bounding box volume.
	
	A related class are the clustered designs, e.g., see Figure~\ref{fig:task2}d created from the layout of Figure~\ref{fig:task2}c. Here we note that the final layout is less dense and multiple thematic clusters appear. Four participants preferred that the larger words were separated and serve as the centers of thematic clusters. In those word clouds, the larger words were spread out and the smaller words that were misplaced or at the periphery between two clusters were brought closer to a certain cluster. Every such design in our study achieved a compactness score of less than 50\%.
	
	The last designs are the mind map designs. In those cases, participants disregarded the initial layout and instead created from scratch a new layout, where small topic bubbles containing few words were spread around the canvas around the central most meaningful word.
	
	We found that participants consistently spent time fixing the reading direction of some word pairings, e.g., the two words ``induced" and ``subgraph" should not be reverted. Additionally, they separated words that were loosely related when they had the same font size and appeared side to side, as they would otherwise appear visually as a word pair rather than two independent words.
	
	\paragraph{Metrics.} The importance of the metrics was evaluated next. The participants were asked which metric, if any, they were interested in when working with the MySemCloud layout. We hypothesized that participants would naturally lean towards compact layouts and realized adjacencies. Seven participants noted they mostly were interested in compactness. As for the semantic metrics, four reported paying attention to the distortion value, and five to realized adjacencies.
	They were then asked how the value of the metrics was correlated with the quality of the layout. Those answers did not align well with the participants' personal design goals. 
	Specifically, eight participants thought that distortion was the most meaningful metric, six thought distortion and realized adjacencies were equally the most relevant and three thought that adjacencies were a better indicator of semantic quality. 
	Compactness was not received as well as the general impression was that it was misleading. Five participants noted that higher compactness lead to worse layouts, and three that it was a secondary goal and only beneficial up to a point.

	\paragraph{MySemCloud.} We hypothesized that our users would enjoy the playful nature of the tool, and appreciate the novelty of semantic word clouds as opposed to the more commonly seen compact non semantic layouts. We also believed that the users would rate highly the ability to interact and improve the visualization over the current best performing layout algorithms. The impressions of MySemCloud as an interactive editor were very positive. Participants valued the simplicity and efficiency of the design (8). Nine participants highlighted the different interaction modes and six the metric guide views. Two participants were interested in the ability to see the underlying data and noted that data exploration was for them a strong use case for MySemCloud. 
	
	\begin{figure}
		\centering
		\includegraphics{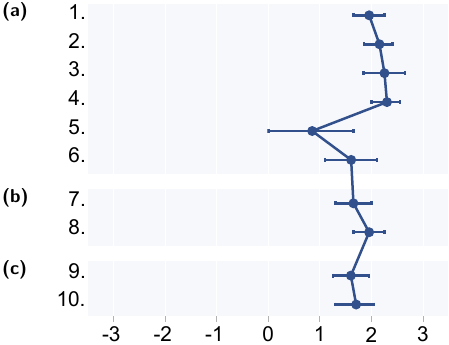}
		\caption{Participants used a seven point Likert scale to rate the following ten statements from \textit{strongly disagree} ($-3$) to \textit{strongly agree} ($3$): 1.) \textit{It was easy to learn MySemCloud (MSC)}, 2.) \textit{It was easy to use MSC}, 3.) \textit{I liked to use MSC}, 4.) \textit{It was fun to use MSC}, 5.) \textit{I felt creative while using MSC}, 6.) \textit{I am satisfied with the result}, 7.) \textit{The metrics were understandable}, 8.) \textit{The metric guides were understandable}, 9.) \textit{MSC represented the data well}, and 10.) \textit{MSC was faithful to the data}; \textbf{\textsf{(a)}} (1.--6.) covers the user experience, \textbf{\textsf{(b)}} (7.--8.) the metrics and \textbf{\textsf{(c)}} (9.--10.) shows that our tool was preferred to the automated layouts.}
		\label{fig:questions}
	\end{figure}
	
	We evaluated our tool using the same questionnaire developed to evaluate ManiWordle~\cite{KohLKS10} and EdWordle~\cite{WangCBZDCS18}, as well an additional set of questions targeted at our system. We can see in Figure~\ref{fig:questions}a that the design of my MySemCloud was found to be very efficient. We can note that creativity was not rated as highly by the users, which is to be expected as we designed the tool to create effective semantic text visualizations as a focus over aesthetics. Moreover, the semantic-enhanced interactions somewhat restrict the edits of the word clouds compared to the free drag-and-drop mode.
	The participants rated highly the quality of the visualization they created using the tool, and thought that the tool itself represented the underlying information of their chosen input text faithfully (see Figure~\ref{fig:questions}c).
	Additionally, while some participants reported issue with the distortion metric, on average when considering all the metrics, the participants had a strong understanding of the optimization goals of the semantic word clouds (see Figure~\ref{fig:questions}b).
	
	Lastly, when comparing their experience of MySemCloud to the non-interactive layout algorithms, 16 participants preferred MySemCloud, with 7 indicating a strong preference.
	On their preference of semantic word clouds over traditional word cloud layouts, 17 participants preferred semantic layouts, two found that it depends on the use case and one participant found that semantic layout presented them with too much information. Additionally, two participants answered that they naturally assumed non-semantic word clouds had a semantically meaningful layout, and thus found them misleading.
	
	\subsection{Limitations}
	
	Interactive word clouds offer more than updating the position of words. For example ManiWordle~\cite{KohLKS10} and EdWordle~\cite{WangCBZDCS18} allow the user to rotate or color words. Such functionality has not yet been implemented in MySemCloud, but could be appealing for users. The focus of MySemCloud so far is the semantic quality of the layout, but it could be worthwhile to study how a system that supports both general aesthetics and semantics is perceived. Such an in-depth system could overwhelm the user, but as some participants in our study suggested, these are possible extensions of MySemCloud.
	
	Additionally, we chose not to allow the user to interact with and modify the underlying data computed by the NLP algorithm, to ensure that it remains faithful to the input text. In the creative part of our user study, however, some participants had complaints about the limitation of the NLP library, e.g., as we often worked with text from mathematical publications, the word ``theorem'' could appear, which the participant considered to not be actually relevant, but the NLP algorithms considered it significant due to its high frequency in the text. Similarly, when dealing with technical vocabulary, stemming can fall short, e.g., the words ``parameterized'' and ``parameter'' being considered two independent words. Thus, the need to remove and add some words from the top-$k$ word list is a natural addition to MySemCloud. While editing the input data might help generate a better initial layout, users will still need to refine the visualization further. Therefore we chose to focus on the more difficult task of providing meaningful information to the user to guide the direct interactions with the layout, while adding data editing modes remains as future work.
	
	With regards to the interactive modes, we noticed that the participants would sometimes move a word on the boundary slightly to trigger some compaction using fill holes; so having a button to trigger the compaction directly would be a natural addition.
	Lastly, the implementation of the bounding boxes sometimes caused visual confusion, and the participants would attempt to bring two words together that would not stay close. This is due to the perception of the bounding box by the user being different from the bounding box used in the algorithm. An implementation similar to EdWordle that considers the bounding box of each letter individually might lessen those issues.

	\section{Conclusion}
	
	In this paper we presented MySemCloud, a novel human-in-the-loop word cloud editor combining the strength of the semantic word cloud layout algorithms with those of interactive word cloud systems. 
	
	We found that users were often dissatisfied with layouts computed by state-of-the-art algorithms as they tended to focus on the wrong semantic relationships, had sometimes undesirable layouts for the largest words, and would misplace several words. The study participants, on average, outperformed the state-of-the-art algorithms on almost all quality metrics with our human-in-the-loop approach. We also found that the focus on compactness provided by previous word cloud editors was detrimental to a visualization of high semantic quality, and the lack of semantic information made the word clouds less interesting.
	
	Overall, we showed that MySemCloud successfully bridges the gap between non-interactive semantic layouts and the existing non-semantic interactive tools. While some users did not believe that they could successfully improve a layout and  were happy to spend a longer amount of time to completely recreate their layout, the large majority of our participants fell between those two extremes and engaged happily with our interactive system. We also found that our tool has a strong potential for data exploration, more so when the users do not have expert knowledge of the input data.
	
	A promising avenue for future work would be to grow the different interactions offered by MySemCloud, most importantly the ability to select which words should or should not be displayed. Further interactive possibilities should be considered carefully as not to overwhelm the user. One can also consider iterating on the current interactive modes. %
	The neighbors-follow mode requires a certain similarity value as a threshold, and it, as well as the fill-holes mode, reactivate forces at a set strength coefficient. These parameters could be set by the user, and would offer flexibility at the cost of making the tool more complex.

	It could also be interesting to study layout methods that take into account user preferences to generate the visualization. Some users are not willing to spend time carefully editing a layout, but still have preferences about which semantic elements should be highlighted. Such a user-centered layout algorithm would combine naturally with our human-in-the-loop fine-tuning system. Lastly, one could also consider extending this approach to handle time-varying data, as the words need to be laid out not only with regards to user preferences but also to enable morphing and optimize stability between subsequent word clouds.

	\acknowledgments{
		We acknowledge support by the Austrian Science Fund (FWF) under grant P31119.}

	\bibliographystyle{abbrv-doi}

	\bibliography{template}
\end{document}